\documentclass[pra,onecolumn,showpacs,preprintnumbers,amsmath,amssymb,times]{revtex4}

\usepackage{psfrag}
\usepackage{epsfig}

\newcommand{\nn}{\nonumber}
\newcommand{\bra}{\langle}
\newcommand{\ve}{\vert}
\newcommand{\ket}{\rangle}

\newcommand{\ee}{\textrm{e}}
\newcommand{\lk}{\left}
\newcommand{\rk}{\right}

\begin{document}


\title[Excitation and Entanglement Transfer Versus Spectral Gap]{Excitation and Entanglement Transfer Versus Spectral Gap}
\author{Michael J. Hartmann}
\email{m.hartmann@imperial.ac.uk}
\author{Moritz E. Reuter}
\author{Martin B. Plenio}
\affiliation{Institute for Mathematical Sciences, Imperial College London,
SW7 2PE, United Kingdom}
\affiliation{QOLS, The Blackett Laboratory, Imperial College London, Prince Consort Road,
SW7 2BW, United Kingdom}
\date{\today}

\begin{abstract}
We consider quantum many body systems as quantum channels and study
the relation between the transfer quality and the size of the spectral
gap between the system's ground and excited states.
In our setup two ancillas are weakly coupled to the quantum
many body system at different sites, and we study the propagation
of an excitation and quantum information from one ancilla to the
other. We observe two different scenarios: a slow, but perfect
transfer if the gap large and a fast, but un-complete transfer
otherwise. We provide a numerical and analytical approach as well as
a simplified physical model explaining our findings.
Our results relate the potential of spin chains acting as quantum channels
to the concept of quantum phase transitions and offer a different
approach to the characterisation of these.
\end{abstract}

\maketitle

---------------------------------------------------------------------------

\section{Introduction} 

Most quantum information processing tasks require at some stage the transfer of quantum states
between two quantum systems such as atoms or ions which are located a different positions in space.
For transfer over long distances, photons sent through optical fibres seem promising.
However the interactions between photons and the stationary systems, e.g. atoms,
are weak and need to be controlled with high precision for transferring
the state onto the photon and vice versa.
Finding alternative methods and carriers is thus of considerable interest,
in particular for transfer over short distances. Here, using condensed matter systems,
e.g. a piece of solid, seems very appealing.
Therefore the possibilities of transferring quantum information with strongly coupled
quantum many body systems such as spin chains have been studied in some detail
in recent years and several scenarios showing close to perfect state transfer have
been found \cite{Bos03,PHE04}. 

One question arises naturally in this context: How do the properties of the employed
many body system relate to the transfer quality and speed?
A key property in this context is whether those systems
feature an energetic gap between their ground state and excited states.
Most interestingly, this gap vanishes at the critical points of quantum phase
transitions \cite{Sachdev1999}, where, at zero temperature, the ground state and
an excited state exchange their roles as a parameter in the
Hamiltonian, such as a magnetic field, is varied.

As with classical phase transitions, quantum phase transitions are usually analysed
in terms of the scaling behaviour of equilibrium properties, where a diverging
correlation length is indicative of a critical point \cite{Has,Sachdev1999,EC05}.
An analogous scaling phenomenon was recently also found for the entanglement
properties of a spin chain in the vicinity of a quantum phase transition
\cite{Osterloh2002,locent,JL05}.
Motivated by these findings and the recent experimental observation
of the Mott quantum phase transition in the well-controlled environment
of an optical lattice \cite{GME+02},
the dynamical entanglement properties of quantum many body systems
undergoing a quantum phase transition are receiving increasing attention. For
example, one recent approach \cite{AUL05} was concerned with the
dynamics of bipartite entanglement in spin chains resulting from
an initial perturbation, while another studied the
entanglement of two spins that are globally coupled to a quantum
critical system \cite{YCW05}. On another level, the Zurek-Kibble
mechanism for classical phase transitions was recently generalised
to its quantum analogue, further deepening our insight into the
dynamics of quantum phase transitions \cite{ZDZ05}.

In this article we study the relation between the size of the
spectral gap of a quantum many body system and its capacity to transfer
quantum information.
Specifically, we study the transfer of
quantum states for two examples of linear chains of interacting
quantum systems. We employ newly developed matrix product state
techniques \cite{Vid03} to simulate numerically the dynamics of
spin chains exhibiting a quantum phase transition. Then we proceed to study a harmonic
chain where we may choose the on-site potential such that the
energy gap above the unique ground state vanishes. The latter of
the two models allows us to obtain a better understanding of the
relevant physics since it permits an analytical study in terms of
master equations and the verification of the validity of the
assumptions inherent in the master equation by numerically
simulating the dynamics of the harmonic chain with up to $1400$
constituents \cite{PHE04}.

We find that the transfer properties crucially depend on the energy gap
between the ground state and the lowest excited states, but does
not significantly dependent on the detailed structure of the
Hamiltonian. 
In particular the transfer The characteristics of the state transfer through
such systems may therefore be used to detect the critical point
experimentally.

%
\section{Spin chains}
\begin{figure}
\begin{center}
\psfrag{1}{\raisebox{-0.12cm}{\hspace{-0.1cm}$1$}}
\psfrag{N}{\raisebox{-0.12cm}{\hspace{-0.12cm}$N$}}
\psfrag{S}{\hspace{-0.2cm}$S$}
\psfrag{R}{$R$}
\psfrag{ms}{\raisebox{0.1cm}{\hspace{-0.1cm}$m_S$}}
\psfrag{mr}{\raisebox{0.1cm}{\hspace{-0.1cm}$m_R$}}
\includegraphics[width=4cm]{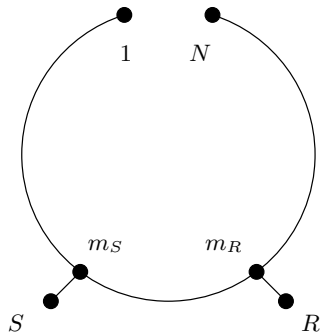}
\caption{\label{topo1} The topology for the spin model considered in the numerical simulations.
$S$ labels the sender and $R$ the receiver ancilla, while $m_S$ and $m_R$ label the spins of the chain
where $S$ and $R$ couple to.} 
\end{center}
\end{figure}

We begin by considering a 1-D chain of spins with nearest neighbour
interactions and open boundary conditions. The Hamiltonian of our
model reads
\begin{equation}\label{hamchain}
H_{\textrm{chain}} = B \sum_{i=1}^N \sigma_i^z + \sum_{i=1}^{N-1} \lk(
J_x \sigma_i^x \sigma_{i+1}^x + J_y \sigma_i^y \sigma_{i+1}^y + J_z \sigma_i^z \sigma_{i+1}^z
\rk) \, ,
\end{equation}
where $N$ is the number of spins, $B$ is an applied magnetic field
and $J_x$, $J_y$ and $J_z$ the interaction between neighboring spins.
Furthermore, two ancillas (named $S$ for ``sender''
and $R$ for ``receiver'') couple to the chain at spins $m_S$ and
$m_R$, which are near the centre of the chain in order to avoid
boundary effects .
The complete Hamiltonian is thus given by 
\begin{equation}\label{hamtot}
H =
H_{\textrm{chain}} + B_a \lk( \sigma_S^z + \sigma_R^z \rk) + J_a
\lk( \sigma_S^x \sigma_{m_S}^x + \sigma_R^x \sigma_{m_R}^x \rk) \, .
\end{equation}
$B_a \ge 0$ is the Zeeman splitting of the ancillas, which might
differ from $B$, and $J_a \ge 0$ is the coupling of the ancillas
to the chain, which is taken to be weak, i.e. $J_a \ll (B, J_x, J_y, J_z)$.
Figure \ref{topo1} shows the topology of the model.

Initially, the chain is assumed to be in the ground state, $\ve 0_{\textrm{chain}} \ket$, of the
Hamiltonian (\ref{hamchain}), while the sender
is spin up and the receiver is spin down. Hence, the initial state
of the total system is
\begin{equation}
\ve \Psi (0) \ket = \ve \uparrow_S, \downarrow_R, 0_{\textrm{chain}} \ket \, .
\end{equation}

We simulate the dynamics of our system numerically, making use of
the recently introduced matrix product states \cite{Vid03}. We use
matrices of dimension $10 \times 10$. To test the accuracy of our simulations,
we verified whether the results where stable with respect to variations of the
matrix dimension and the size of the timesteps. Furthermore, we tested whether
the energy of the total system was conserved. Since the matrix product approximation
can only be efficient if the considered system obeys a "entropy and area law"
\cite{PED+}, which is not necessarily true at quantum critical points, our
simulations consider only parameters near, but not exactly on the
critical point.

\begin{figure}
\begin{center}
\psfrag{t}{\raisebox{-0.4cm}{$t$}}
\psfrag{1}{\hspace{-0.26cm}\tiny{$1.0$}}
\psfrag{0.1}{\hspace{-0.2cm}\tiny{$0.0$}}
\psfrag{0.2}{\hspace{-0.2cm}\tiny{$0.2$}}
\psfrag{0.4}{\hspace{-0.2cm}\tiny{$0.4$}}
\psfrag{0.6}{\hspace{-0.2cm}\tiny{$0.6$}}
\psfrag{0.8}{\hspace{-0.2cm}\tiny{$0.8$}}
\psfrag{0}{\raisebox{-0.1cm}{\tiny{$0$}}}
\psfrag{500}{\raisebox{-0.1cm}{\tiny{$ $}}}
\psfrag{1000}{\raisebox{-0.2cm}{\hspace{-0.2cm}\tiny{$5 \times 10^4$}}}
\psfrag{1500}{\raisebox{-0.1cm}{\tiny{$ $}}}
\psfrag{2000}{\raisebox{-0.2cm}{\hspace{-0.25cm}\tiny{$10 \times 10^4$}}}
\psfrag{2500}{\raisebox{-0.1cm}{\tiny{$ $}}}
\psfrag{3000}{\raisebox{-0.2cm}{\hspace{-0.25cm}\tiny{$15 \times 10^4$}}}
\includegraphics[width=10cm]{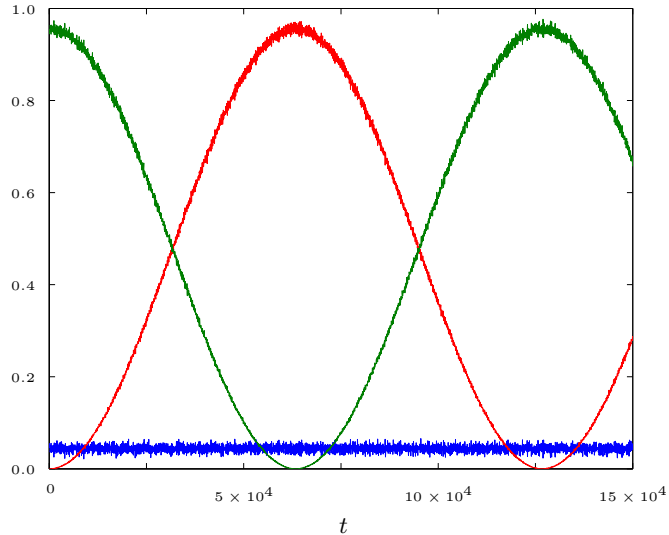}
\caption{\label{simulnonres} $P(\downarrow_S \downarrow_R)(t)$ (blue), $P(\uparrow_S \downarrow_R)(t)$
(green) and $P(\downarrow_S \uparrow_R)(t)$ (red) for $B = 1$, $J_x = 0.3$, $J_y = J_z = 0$, $B_a = 0.64$ and $J_a = 0.05$
as given by the simulation for the open boundary model with $N = 100$ spins.
$S$ couples to spin $45$ and $R$ to spin $55$.}
\end{center}
\end{figure}

Figure \ref{simulnonres} shows the probability  $P(\uparrow_S
\downarrow_R)$ that ``sender'' $S$ is in its excited state $\ve
\uparrow_S \ket$ and the ``receiver'' $R$ in its ground state $\ve
\downarrow_R \ket$, together with $P(\downarrow_S \uparrow_R)$ and $P(\downarrow_S \downarrow_R)$ for a model with $N = 100$, $m_S = 45$, $m_R = 55$, $B =
1$, $J_x = 0.3$, $J_y = J_z = 0$, $B_a = 0.64$ and $J_a = 0.05$. $P(\uparrow_S \uparrow_R)$ is always less than
$10^{-4}$.
%
%
The plots show that the excitation that was initially located in $S$ oscillates back and forth between
$S$ and $R$.

\begin{figure}
\begin{center}
\psfrag{t}{\raisebox{-0.4cm}{$t$}}
\psfrag{1}{\hspace{-0.26cm}\tiny{$1.0$}}
\psfrag{0.1}{\hspace{-0.2cm}\tiny{$0.0$}}
\psfrag{0.2}{\hspace{-0.2cm}\tiny{$0.2$}}
\psfrag{0.4}{\hspace{-0.2cm}\tiny{$0.4$}}
\psfrag{0.6}{\hspace{-0.2cm}\tiny{$0.6$}}
\psfrag{0.8}{\hspace{-0.2cm}\tiny{$0.8$}}
\psfrag{0}{\raisebox{-0.1cm}{\tiny{$0$}}}
\psfrag{200}{\raisebox{-0.1cm}{\hspace{0.0cm}\tiny{$200$}}}
\psfrag{400}{\raisebox{-0.1cm}{\hspace{0.0cm}\tiny{$400$}}}
\psfrag{600}{\raisebox{-0.1cm}{\hspace{0.0cm}\tiny{$600$}}}
\psfrag{800}{\raisebox{-0.1cm}{\hspace{0.0cm}\tiny{$800$}}}
\psfrag{1000}{\raisebox{-0.1cm}{\hspace{0.0cm}\tiny{$1000$}}}
\includegraphics[width=10cm]{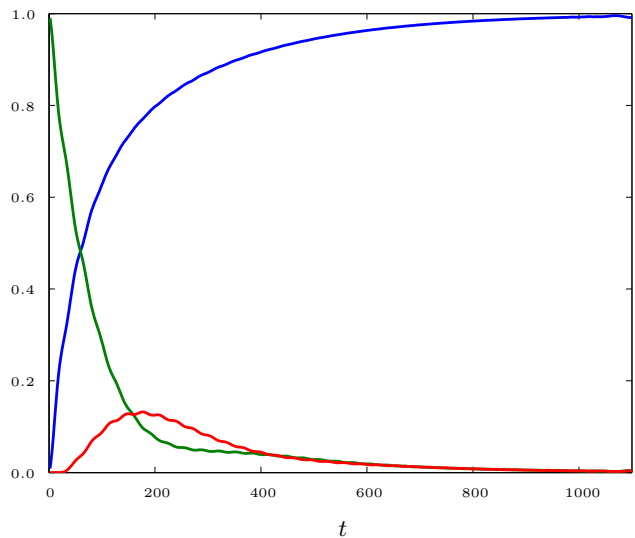}
\caption{\label{simulres} $P(\downarrow_S \downarrow_R)(t)$ (blue), $P(\uparrow_S \downarrow_R)(t)$ (green)
and $P(\downarrow_S \uparrow_R)(t)$ (red) for $B = 1$, $J_x = 0.3$, $J_y = J_z = 0$, $B_a = 0.8$ and
$J_a = 0.05$ as given by the simulation for the open boundary model with $N = 600$ spins.
$S$ couples to spin $295$ and $R$ to spin $305$.}
\end{center}
\end{figure}

Figure \ref{simulres} shows $P(\uparrow_S \downarrow_R)$, $P(\downarrow_S \uparrow_R)$
and $P(\downarrow_S \downarrow_R)$ for
a model with $N = 600$, $m_S = 295$, $m_R = 305$, $B = 1$, $J_x = 0.3$, $J_y = J_z = 0$, $B_a = 0.8$ and $J_a = 0.05$.
Again, $P(\uparrow_S \uparrow_R)$ is always less than
$10^{-4}$.
%
%
For these parameters, the excitation is not fully transferred to
$R$, contrary to figure \ref{simulnonres}. Both, $S$ and $R$ relax
to their ground states with the excitation only being partially
and temporarily transferred to $R$, even for close-lying spins.
Note that the parameters chosen in figures \ref{simulnonres} and \ref{simulres}
are the same except for $B_a$ which in figure \ref{simulres} is significantly larger
than in figure \ref{simulnonres}.

\begin{figure}
\begin{center}
\psfrag{t}{\raisebox{-0.4cm}{$t$}}
\psfrag{1}{\hspace{-0.26cm}\tiny{$1.0$}}
\psfrag{0.1}{\hspace{-0.2cm}\tiny{$0.0$}}
\psfrag{0.2}{\hspace{-0.2cm}\tiny{$0.2$}}
\psfrag{0.4}{\hspace{-0.2cm}\tiny{$0.4$}}
\psfrag{0.6}{\hspace{-0.2cm}\tiny{$0.6$}}
\psfrag{0.8}{\hspace{-0.2cm}\tiny{$0.8$}}
\psfrag{0}{\raisebox{-0.1cm}{\tiny{$0$}}}
\psfrag{2000}{\raisebox{-0.1cm}{\hspace{-0.0cm}\tiny{$2000$}}}
\psfrag{4000}{\raisebox{-0.1cm}{\hspace{-0.0cm}\tiny{$4000$}}}
\psfrag{6000}{\raisebox{-0.1cm}{\hspace{-0.0cm}\tiny{$6000$}}}
\psfrag{8000}{\raisebox{-0.1cm}{\hspace{-0.0cm}\tiny{$8000$}}}
\psfrag{10000}{\raisebox{-0.1cm}{\hspace{-0.0cm}\tiny{$10000$}}}
\includegraphics[width=10cm]{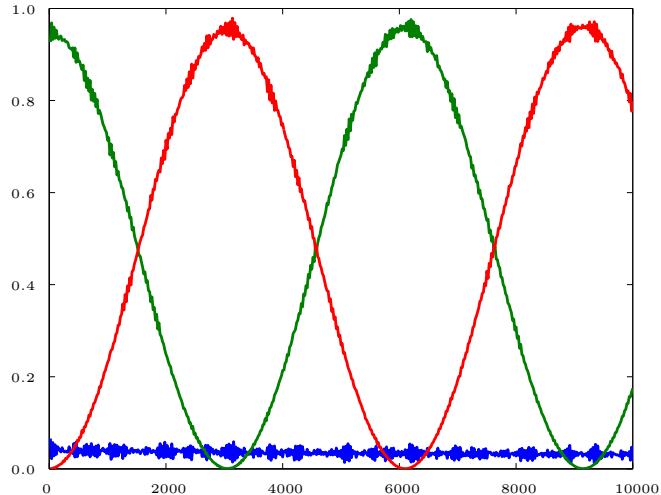}
\caption{\label{simulnonres1} $P(\downarrow_S \downarrow_R)(t)$ (blue),
$P(\uparrow_S \downarrow_R)(t)$ (green)
and $P(\downarrow_S \uparrow_R)(t)$ (red)
for $B = 1$, $J_x = 0.5$, $J_y = 0.2$, $J_z = 0.1$, $B_a = 0.04$ and $J_a = 0.05$
as given by the simulation for the open boundary model with $N = 100$ spins.
$S$ couples to spin $45$ and $R$ to spin $55$.}
\end{center}
\end{figure}

\begin{figure}
\begin{center}
\psfrag{t}{\raisebox{-0.4cm}{$t$}}
\psfrag{1}{\hspace{-0.26cm}\tiny{$1.0$}}
\psfrag{0.1}{\hspace{-0.2cm}\tiny{$0.0$}}
\psfrag{0.2}{\hspace{-0.2cm}\tiny{$0.2$}}
\psfrag{0.4}{\hspace{-0.2cm}\tiny{$0.4$}}
\psfrag{0.6}{\hspace{-0.2cm}\tiny{$0.6$}}
\psfrag{0.8}{\hspace{-0.2cm}\tiny{$0.8$}}
\psfrag{0}{\raisebox{-0.1cm}{\tiny{$0$}}}
\psfrag{100}{\raisebox{-0.1cm}{\tiny{$200$}}}
\psfrag{200}{\raisebox{-0.1cm}{\tiny{$400$}}}
\psfrag{300}{\raisebox{-0.1cm}{\tiny{$600$}}}
\includegraphics[width=10cm]{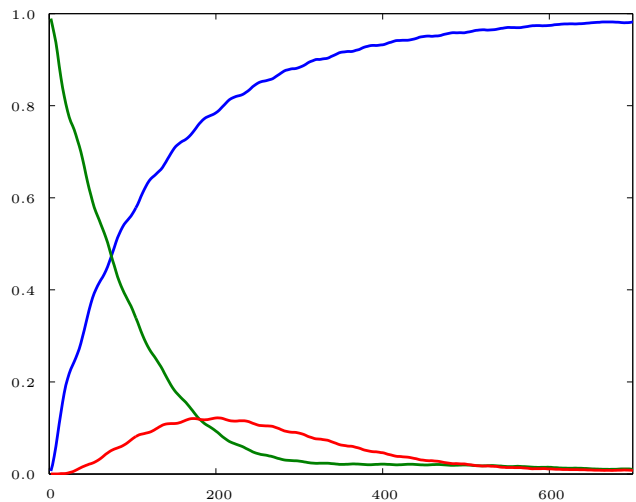}
\caption{\label{simulres1} $P(\downarrow_S \downarrow_R)(t)$ (blue),
$P(\uparrow_S \downarrow_R)(t)$ (green)
and $P(\downarrow_S \uparrow_R)(t)$ (blue)
for $B = 1$, $J_x = 0.3$, $J_y = 0.2$, $J_z = 0.1$, $B_a = 0.2$ and $J_a = 0.05$
as given by the simulation for the open boundary model with $N = 600$ spins.
$S$ couples to spin $295$ and $R$ to spin $305$.}
\end{center}
\end{figure}

The two observed scenarios are rather generic. To demonstrate this,
we have done the same simulations for different parameters, i.e. for a XXZ-model.
The results, shown in figures \ref{simulnonres1} and \ref{simulres1},
clearly agree with our findings for the previous coupling parameters.
Again $B_a$ in figure \ref{simulres1} is significantly larger
than in figure \ref{simulnonres1}, while all other parameters are equal.

%
\section{Heuristic physical picture}
\begin{figure}
\begin{center}
\psfrag{S}{\raisebox{-0.2cm}{\hspace{-0.1cm}$S$}}
\psfrag{R}{\raisebox{-0.2cm}{\hspace{-0.1cm}$R$}}
\psfrag{Chain}{\raisebox{-0.2cm}{\hspace{-0.1cm}chain}}
\psfrag{Re0}{\raisebox{0.1cm}{\hspace{-0.6cm}$ $}}
\psfrag{Ren0}{\raisebox{0.1cm}{\hspace{-0.6cm}$ $}}
\includegraphics[width=4cm]{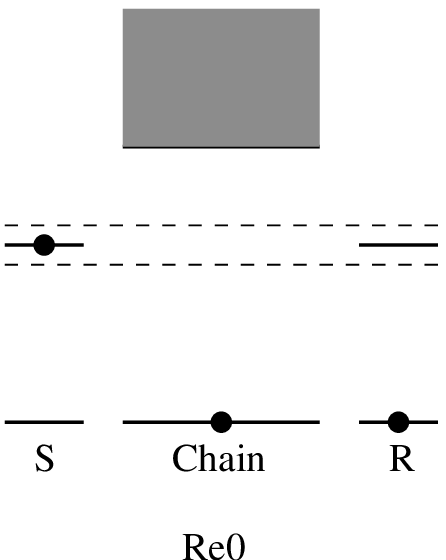}
\hspace{2cm}
\includegraphics[width=4cm]{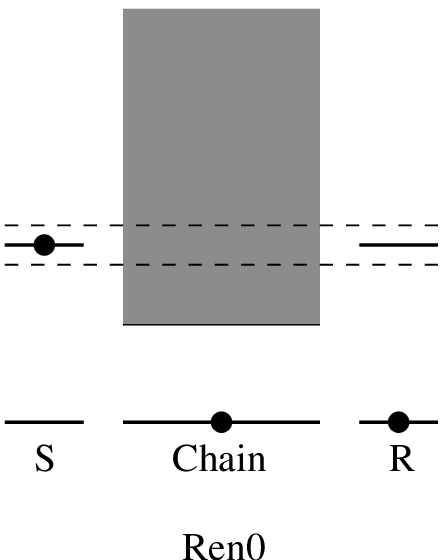}
\caption{\label{energycon} Sketch of the energy levels of the system. The dots indicate the occupations
of the initial state. For this initial state only the energy levels between the two horizontal
dashed lines are accessible, resulting in almost perfect transfer for the left scenario
and damping for the right one.}
\end{center}
\end{figure}

The dramatic difference between the almost perfect transfer scenarios in figures
\ref{simulnonres} and \ref{simulnonres1} and the damped scenario in figures \ref{simulres} and
\ref{simulres1} has a simple physical explanation.
The dynamics we have simulated is given by the Schr\"odinger equation containing the
Hamiltonian (\ref{hamtot}). As a consequence, all moments of the Hamiltonian are conserved,
\begin{equation}
\bra \Psi (t) \ve H^n \ve \Psi (t) \ket = \bra \Psi (0) \ve H^n \ve \Psi (0) \ket = \textrm{const}
\quad \textrm{for any integer} \enspace n
\end{equation}
%
The initial state $\ve \Psi (0) \ket$ is not an eigenstate of $H$ as given by (\ref{hamtot}), hence
\begin{equation}
\bra \Psi (0) \ve H \ve \Psi (0) \ket = \sum_E |\bra E | \Psi (0) \ket|^2 \, E \, ,
\end{equation}
where $E$ and $| E \ket$ are the eigenvalues and eigenstates of $H$.
However since a probability distribution is entirely determined by all moments,
$\bra \Psi (t) \ve H^n \ve \Psi (t) \ket = \enspace$const for all $n$ implies
$|\bra E | \Psi (t) \ket|^2 = |\bra E | \Psi (0) \ket|^2$ for all $| E \ket$.
In other words the whole probability distribution given by the $|\bra E | \Psi (0) \ket|^2$
is conserved. In our case, it's variance is
\begin{equation}
\sqrt{\bra \Psi (t) \ve H^2 \ve \Psi (t) \ket - \bra \Psi (t) \ve H \ve \Psi (t) \ket^2} = J_a \, .
\end{equation}
For the dynamics this means that only those states with an energy expectation value
$\overline{E}$ in the range $\bra \Psi (0) \ve H \ve \Psi (0) \ket - 2 J_a <
\overline{E} < \bra \Psi (0) \ve H \ve \Psi (0) \ket + 2 J_a$ are accessible.
Figure \ref{energycon} sketches the energy levels of the system we consider.
$S$ and $R$ are depicted as two level systems, while for the chain there is a unique ground state
and a quasi continuous band of excited states sketched as the gray area.
The dots indicate the initial occupations.
The energy range which is accessible for the considered initial state lies between the two
horizontal dashed lines. If the spectral gap is larger than the Zeeman splitting of the ancillas
(left plot), there is no accessible excited state of the chain and hence no excitations get lost
into the chain, which in turn implies the excitation will be almost completely transferred to $R$.
If however the spectral gap is smaller than the Zeeman splitting of the ancillas,
there are accessible excited states in the chain and excitation and hence quantum information get lost.  

In order to obtain a more rigorous justification of this simple picture
and to underline the generality of our findings, we now turn to a different model for
the chain which also features an adjustable energy gap above its
unique ground state. 

%
\section{Harmonic chain}
\begin{figure}
\begin{center}
\psfrag{1}{$1$}
\psfrag{N}{\hspace{-0.14cm}$N$}
\psfrag{S}{\hspace{-0.16cm}$S$}
\psfrag{R}{$R$}
\psfrag{ms}{$m_S$}
\psfrag{mr}{\hspace{-0.16cm}$m_R$}
\includegraphics[width=4cm]{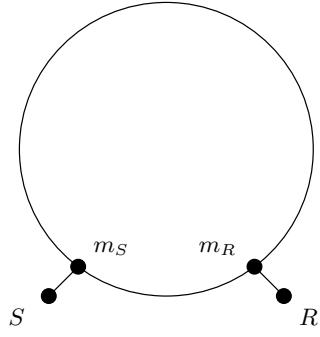}
\caption{\label{topo2} The topology for the harmonic chain model considered in the analytical approach.
$S$ labels the sender and $R$ the receiver ancilla, while $m_S$ and $m_R$ label the oscillators of the chain
where $S$ and $R$ couple to.} 
\end{center}
\end{figure}

We consider a harmonic chain with periodic boundary conditions (see fig. \ref{topo2})
described by
\begin{equation} \label{harmchain}
H_{\textrm{chain}} = \frac{1}{2} \sum_{j=1}^N \lk( p_j^2 + \Omega^2 (q_j - q_{j+1})^2 +
\Omega_0^2 q_j^2 \rk)
\end{equation}
with the $p_j$ being the momenta and the $q_j$ the positions
($q_{N+1} = q_1$). In this case, the two ancillas are harmonic
oscillators that couple to oscillators $m_S$ and $m_R$ of the
chain. The complete Hamiltonian now reads
\begin{eqnarray} \label{harmtot}
H_{\textrm{tot}} & = & H_{\textrm{chain}} + H_{\textrm{ancillas}} + H_I \\
H_{\textrm{ancillas}} & = & \frac{1}{2} \lk( p_S^2 + \omega^2 q_S^2 + p_R^2 + \omega^2 q_R^2 \rk) \\
H_I & = & J_a (q_S q_{m_S} + q_R q_{m_R}) \, .  
\end{eqnarray}
Since we are only interested in the time evolution of the ancillas, we derive a master equation
for the dynamics of their reduced density matrix $\rho (t)$.
For weak coupling $J_a \ll (\Omega, \Omega_0)$, its equation of motion is given by
\begin{equation}\label{nakajima}
\frac{d\sigma}{dt} = - \int_0^t ds \, \textrm{Tr}_{\textrm{chain}}
\lk\{ \lk[ H_I(t), \lk[ H_I(s), \ve 0 \ket \bra 0 \ve \otimes
\sigma(s) \rk] \rk] \rk\} \, ,
\end{equation}
where $\sigma(t)$ and $H_I(t)$ are the density matrix of the ancillas and
the interaction between ancillas and chain in the interaction
picture, respectively: $H = H_0 + H_I$ with
$H_0 = H_{\textrm{chain}} + H_{\textrm{ancillas}}$,
$H_I(t) = \exp (i H_0 t) H_I \exp (- i H_0 t)$ and $\sigma(t) =
\exp (i H_0 t) \rho \exp (- i H_0 t)$. $\textrm{Tr}_{\textrm{chain}}$
is the trace over the degrees of freedom of the chain and $\ve 0 \ket$
denotes the ground state of the chain (\ref{harmchain}).
The right hand side of eq. (\ref{nakajima}) is an expansion in the
coupling strength $J_a$ up to second order,
which is a good approximation if the integral approaches a
constant value for $t > t^{\star}$, where $J_a t^{\star} \ll 1$.
Since $\sigma$ only changes significantly on time scales
$t \sim J_a^{-1} \gg t^{\star}$, the approximation
$\sigma(s) \approx \sigma(t)$ can be used. Performing the
trace on the rhs of (\ref{nakajima}) yields
\begin{eqnarray} \label{masterform}
\frac{d\sigma}{dt} = - J_a^2 \sum_{j, l = S, R} &&\lk(
- i \lk(Y_1 + (Y_0 - Y_1)\delta_{jl}\rk) [a_j a_l^{\dagger}, \sigma] \rk. \nn \\
& & \enspace + \lk. \lk(X_1 + (X_0 - X_1)\delta_{jl}\rk) \lk( \{ a_j a_l^{\dagger}, \sigma \}
- 2 ( a_j \sigma a_l^{\dagger} )\rk) \rk) \, ,
\end{eqnarray}
where $a_S$ and $a_R$ are the annihilation operators of $S$ and
$R$, respectively: $q_j = (a_j + a_j^{\dagger})/\sqrt{2 \omega}$
and $p_j = - i (a_j - a_j^{\dagger}) \sqrt{\omega / 2}$ for $j = S, R$.
$[ \cdot, \cdot ]$ and $\{ \cdot, \cdot \}$ denote commutators and
anti-commutators. On the rhs of the above equation, we neglected
terms which contain two annihilation or two creation operators since they
oscillate at high frequencies.
The validity of this approximation can later be confirmed from the exact
numerical solution. The coefficients read
\begin{eqnarray}
X_0 = \textrm{Re}(C_{m_S m_S}^+ + C_{m_S m_S}^-) / 2 \omega \, ,\\
X_1 = \textrm{Re}(C_{m_S m_R}^+ + C_{m_S m_R}^-) / 2 \omega \, ,\\
Y_0 = \textrm{Im}(C_{m_S m_S}^+ + C_{m_S m_S}^-) / 2 \omega \quad \textrm{and} \\
Y_1 = \textrm{Im}(C_{m_S m_R}^+ + C_{m_S m_R}^-) / 2 \omega
\end{eqnarray}
with $C_{kl}^{\pm}$ given by
\begin{equation}
C_{kl}^{\pm} (t) = \int_0^t ds \, \bra 0 \ve q_k (t) q_l (s) \ve 0 \ket \,
\ee^{\pm i \omega (t-s)} \, ,
\end{equation}
where $k, l = m_S, m_R$.
Due to the symmetries of the model, the $C_{kl}^{\pm}$ only depend
on $|k - l|$, implying $C_{m_S m_S}^{\pm} = C_{m_R m_R}^{\pm}$ and
$C_{m_S m_R}^{\pm} = C_{m_R m_S}^{\pm}$.
Eq. (\ref{masterform}) is a good approximation whenever
\begin{equation} \label{casym}
C_{kl}^{\pm} (t) \approx \overline{C}_{kl}^{\pm} = \textrm{const.} \quad \textrm{for} \quad t \ll J_a^{-1} \, .
\end{equation}
Since the $C_{kl}^{\pm} (t)$ do not depend
on $J_a$ themselves, there is always a sufficiently small $J_a$
such that (\ref{casym}) holds, provided $\lim_{t \to \infty} C_{kl}^{\pm}
(t)$ exists.

The harmonic chain can be diagonalised via a Fourier transform \cite{PHE04}.
In the limit of an infinitely long chain, $N \rightarrow \infty$,
its dispersion relation is
\begin{equation}
\omega_k^2 = 4 \Omega^2 \sin^2 \frac{k}{2} + \Omega_0^2 \, , \quad - \pi < k < \pi \, ,
\end{equation}
and the correlation functions read
\begin{equation}
\bra 0 \ve q_j (t) q_l (s) \ve 0 \ket =
\frac{1}{2 \pi} \int_0^{\pi} dk \, \omega_k^{-1} \, \cos((j - l) k) \, \ee^{- i \omega_k (t - s)} \, .
\end{equation}
These expressions show, that indeed all $\lim_{t \to
\infty} C_{kl}^{\pm} (t)$ exist except for the case where
$\omega = \Omega_0 = 0$.
As in master equations for system bath
models, we now insert the asymptotic expressions
\begin{equation} \label{mastercoeff}
\overline{C}_{kl}^{\pm} = \lim_{t \to \infty} C_{kl}^{\pm} (t) = 
\int_0^{\infty} ds \, \bra 0 \ve q_k (t) q_l (s) \ve 0 \ket \, \ee^{\pm i \omega (t-s)}
\end{equation}
into eq. (\ref{masterform}). This replacement assumes that all
internal dynamics of the chain happens on much shorter time scales
than the dynamics caused by the interaction of the ancillas with
the chain. Furthermore, it does not treat the initial evolution for
short times with full accuracy since $\lim_{t \to 0} C_{kl}^{\pm}
(t) = 0 (\not= \overline{C}_{kl}^{\pm})$. The obtained master
equation is thus valid in a regime where the couplings $J_a$ are
weak enough such that the time it takes for an excitation to travel from
$S$ to $R$ is completely determined by $J_a$, i.e. by the time it
takes to be transferred into and from the chain. Consequently, the
speed of sound of the chain is no longer resolved, and differences
in the distance between $S$ and $R$ do not matter.

From eq. (\ref{masterform}) we find the following solution for the
expectation values of the occupation numbers of $S$ and $R$,
$n_S = \textrm{Tr}(a_S^{\dagger} a_S \sigma)$ and $n_R = \textrm{Tr}(a_R^{\dagger} a_R \sigma)$:
\begin{equation} \label{mastersol}
\lk.
\begin{array}{r}
n_S (t)\\
n_R (t)
\end{array}
\rk\} = \lk( A_+ \cosh(2 J_a^2 x_1 t) \pm A_-\cos(2 J_a^2 y_1 t) \rk) \, \exp\lk(- 2 J_a^2 x_0 t \rk)
\end{equation}
Here, $A_+ = \frac{n_S(0) + n_R(0)}{2}$, $A_- = \frac{n_S(0) - n_R(0)}{2}$, $x_0 = \lim_{t \to \infty} X_0$, $x_1 = \lim_{t \to \infty} X_1$ and $y_1 = \lim_{t \to \infty} Y_1$.
Note that $x_0 > 0$ and $x_0 > |x_1|$.

\begin{figure}
\begin{center}
\psfrag{t}{\raisebox{-0.4cm}{$t$}}
\psfrag{1}{\hspace{-0.3cm}\tiny{$1.0$}}
\psfrag{0.1}{\hspace{0.1cm}\tiny{$0.0$}}
\psfrag{0.2}{\hspace{0.1cm}\tiny{$0.2$}}
\psfrag{0.4}{\hspace{0.1cm}\tiny{$0.4$}}
\psfrag{0.6}{\hspace{0.1cm}\tiny{$0.6$}}
\psfrag{0.8}{\hspace{0.1cm}\tiny{$0.8$}}
\psfrag{0}{\raisebox{0.1cm}{\tiny{$0$}}}
\psfrag{200}{\raisebox{-0.0cm}{\hspace{0.1cm}\tiny{$2 \times 10^4$}}}
\psfrag{400}{\raisebox{-0.0cm}{\hspace{0.1cm}\tiny{$ $}}}
\psfrag{600}{\raisebox{-0.0cm}{\hspace{0.1cm}\tiny{$6 \times 10^4$}}}
\psfrag{800}{\raisebox{-0.0cm}{\hspace{0.1cm}\tiny{$ $}}}
\psfrag{1000}{\raisebox{-0.0cm}{\hspace{0.1cm}\tiny{$10^5$}}}
\includegraphics[width=10cm]{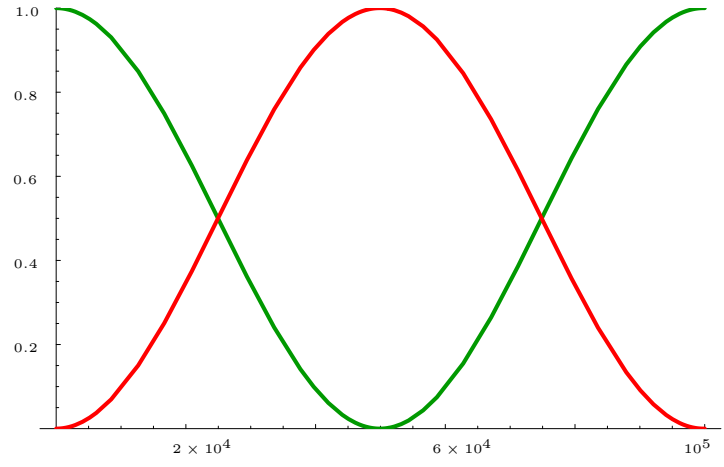}
\caption{\label{mastersolplot1} Solution (\ref{mastersolnonres}) for a harmonic chain and ancillas
with $\Omega = 1$, $\omega = 0.5$, $J_a = 0.05$, $|m_S - m_R| = 9$ and $\Omega_0 = 0.7$}
\end{center}
\end{figure}

Inserting the correlation functions into eq. (\ref{mastercoeff}) and
using the relation
\begin{equation} 
\int_{- \infty}^{\infty} dx \, f(x) \int_0^{\infty} d\tau \, \ee^{- i x \tau} =
\pi f(0)- i \mathcal{P} \int_{- \infty}^{\infty} dx \, \frac{f(x)}{x} \, ,
\end{equation}
where $\mathcal{P}$ denotes the principal value of the subsequent
integral, one sees that $x_0$ and $x_1$ are only non-zero if
$\omega \ge \omega_k$ for at least one mode $k$, that is if our
initial state is in resonance with (i.e. has the same energy
expectation value as) states where both ancillas are in their
ground states and the chain is in one of its lowest-lying excited
states (c.f. figure \ref{energycon}). The dispersion relation shows that this only happens for
$\omega \ge \Omega_0$. As in figures \ref{simulnonres} and \ref{simulres} or
\ref{simulnonres1} and \ref{simulres1}, we thus observe two different scenarios:

If $\omega < \Omega_0$, and therefore $x_0 = x_1 = 0$,
the excitation that is initially in $S$ oscillates back and
forth between $S$ and $R$ at a frequency $2 J_a^2 y_1$, i.e.
\begin{equation} \label{mastersolnonres}
\lk.
\begin{array}{r}
n_S (t)\\
n_R (t)
\end{array}
\rk\} = A_+ \pm A_-\cos(2 J_a^2 y_1 t) \, .
\end{equation}
Note in particular that the excitation is entirely transferred to $R$ at times
$t_n = n (\pi / J_a^2 y_1); \: n = 1, 2, \dots$ . The solution (\ref{mastersolnonres})
is plotted in figure \ref{mastersolplot1} for a harmonic chain and ancillas with
$\Omega = 1$, $\omega = 0.5$, $J_a = 0.05$, $|m_S - m_R| = 9$ and $\Omega_0 = 0.7$.

\begin{figure}
\begin{center}
\psfrag{o}{\small{$\Omega_0$}}
\psfrag{f}{\small{$2 \, J_a^2 \, y_1$}}
\psfrag{6}{\hspace{-0.06cm}\raisebox{-0.04cm}{\tiny{$0.6$}}}
\psfrag{7}{\hspace{-0.06cm}\raisebox{-0.04cm}{\tiny{$0.7$}}}
\psfrag{8}{\hspace{-0.06cm}\raisebox{-0.04cm}{\tiny{$0.8$}}}
\psfrag{9}{\hspace{-0.06cm}\raisebox{-0.04cm}{\tiny{$0.9$}}}
\psfrag{0.0002}{\tiny{$ $}}
\psfrag{0.0004}{\hspace{0.4cm}\tiny{$0.0004$}}
\psfrag{0.0006}{\tiny{$ $}}
\psfrag{0.0008}{\hspace{0.4cm}\tiny{$0.0008$}}
\psfrag{0.001}{\tiny{$ $}}
\psfrag{0.0012}{\hspace{0.4cm}\tiny{$0.0012$}}
\psfrag{0.0014}{\tiny{$ $}}
\includegraphics[width=10cm]{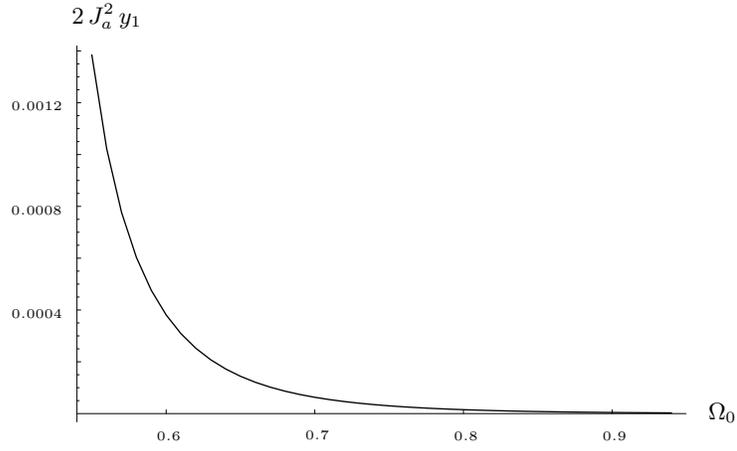}
\caption{\label{offresplot} Frequencies $2 J_a^2 y_1$ of the excitation's oscillation between
$S$ and $R$ for $\Omega = 1$, $J_a = 0.05$ and $\omega = 0.5$. The transfer speed increases
as $\Omega_0 - \omega \rightarrow 0$.}
\end{center}
\end{figure}

Figure \ref{offresplot} shows the frequencies $2 J_a^2 y_1$
of the excitation's oscillations between $S$ and $R$ for cases where $\omega < \Omega_0$,
for $\Omega = 1$, $J_a = 0.05$ and $\omega = 0.35$ as a function of $\Omega_0$.
As $\Omega_0 - \omega$ decreases, the transfer becomes faster and the oscillation frequency
increases.
%
%

\begin{figure}
\begin{center}
\psfrag{t}{\raisebox{-0.4cm}{$t$}}
\psfrag{1}{\hspace{-0.3cm}\tiny{$1.0$}}
\psfrag{0.1}{\hspace{0.1cm}\tiny{$0.0$}}
\psfrag{0.2}{\hspace{0.1cm}\tiny{$0.2$}}
\psfrag{0.4}{\hspace{0.1cm}\tiny{$0.4$}}
\psfrag{0.6}{\hspace{0.1cm}\tiny{$0.6$}}
\psfrag{0.8}{\hspace{0.1cm}\tiny{$0.8$}}
\psfrag{500}{\raisebox{-0.0cm}{\hspace{0.1cm}\tiny{$500$}}}
\psfrag{1000}{\raisebox{-0.0cm}{\hspace{0.1cm}\tiny{$1000$}}}
\psfrag{1500}{\raisebox{-0.0cm}{\hspace{0.1cm}\tiny{$1500$}}}
\psfrag{2000}{\raisebox{-0.0cm}{\hspace{0.1cm}\tiny{$2000$}}}
\includegraphics[width=10cm]{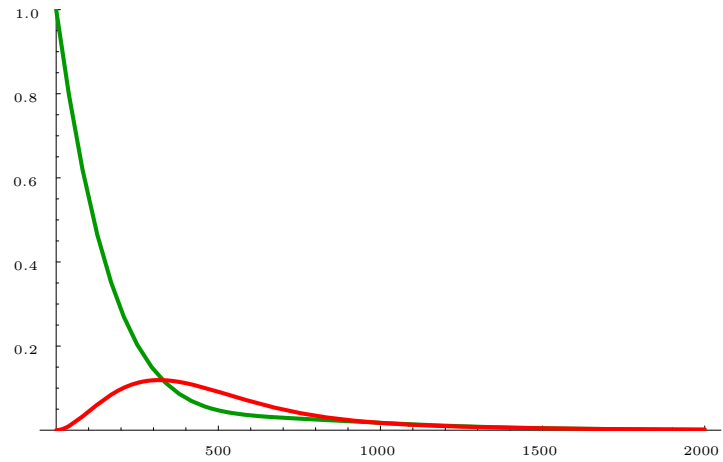}
\caption{\label{mastersolplot2} The solution (\ref{mastersol}) for a
harmonic chain and ancillas with $\Omega = 1$, $\omega = 0.5$, $J_a = 0.05$,
$|m_S - m_R| = 9$ and $\Omega_0 = 0.2$.}
\end{center}
\end{figure}

If, on the other hand, $\omega \ge \Omega_0$, the chain acts
similarly to a bath. Here, $x_0 \not= 0$, $x_1 \not= 0$, and both
ancillas relax into their ground state transferring their energy
into the chain. During this process, however, a fraction of the
energy initially located in $S$ appears momentarily in $R$
before it is finally damped into the chain.
The maximal excitation of the receiver throughout the entire evolution
depends on the distance $|m_S - m_R|$. For a given initial energy in $S$,
a narrow range of the excitation spectrum is relevant for the dynamics.
The relation of the wavelength of these excitations to the distance
$|m_S - m_R|$ determines the maximal transferred fraction of the excitation.
Figure \ref{mastersolplot2} shows the solution \ref{mastersol} for a
harmonic chain and ancillas with $\Omega = 1$, $\omega = 0.5$, $J_a = 0.05$,
$|m_S - m_R| = 9$ and $\Omega_0 = 0.2$.

One might try to derive the same type of master equation for the
spin chain (\ref{hamchain}). However, for finite distances no
exact expression for the time dependent correlation functions is
known \cite{McC71}. This is due to the fact that the subspaces of
odd and of even number of fermions cannot be diagonalised
simultaneously. An attempt of an approximation restricted to only
one subspace led to reasonable results for some parameter values but
occasionally produced unphysical solutions which
grew exponentially in time. Therefore, such a master equation approach
cannot be considered reliable for our spin chains and was avoided.

To confirm the validity of the master equation approach for the
harmonic chain, we compared it to results of a numerical
simulation of a chain with 1400 oscillators.
Since the Hamiltonians of harmonic oscillators and
harmonic chains are quadratic in the position and momentum
operators, Gaussian states (states with a Gaussian Wigner
function) remain Gaussian throughout the time evolution. For these
states the complete dynamics can thus be obtained by only considering
the evolution of the covariance matrix (see \cite{PHE04} for details).
We found good agreement between our analytical and numerical solutions,
with the relative errors being less than $5\%$.

\section{Quantum information transfer}

\begin{figure} \begin{center}
\psfrag{S}{\raisebox{-0.2cm}{\hspace{-0.2cm}$S$}}
\psfrag{R}{\raisebox{-0.2cm}{$R$}}
\psfrag{ms}{$m_S$}
\psfrag{mr}{\hspace{-0.2cm}$m_R$}
\psfrag{C}{\raisebox{-0.2cm}{\hspace{-0.1cm}$C$}}
\includegraphics[width=4cm]{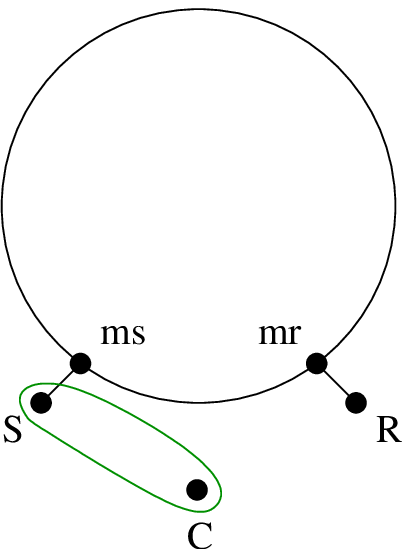}
\hspace{2cm}
\includegraphics[width=4cm]{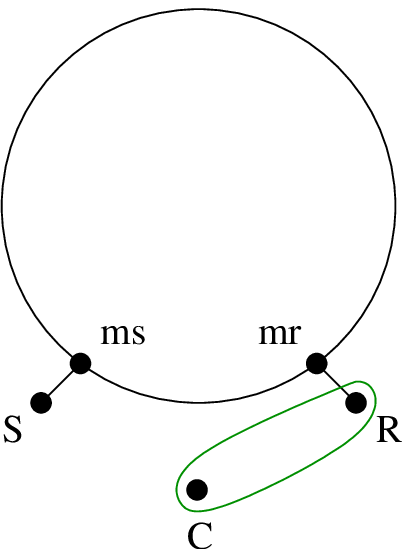}
\caption{\label{entfig} The reference spin $C$ is initially maximally entangled with the sender $S$,
left plot.
If an excitation, which was initially in $S$ gets perfectly transferred to $R$, the entanglement
will then be shared between $R$ and $C$, right plot.}
\end{center} \end{figure}

The observed effects may also be formulated in quantum information language.
In this way, one obtains statements on the average
fidelity achieved for arbitrary input states
(subspace fidelity) or the transfer of
entanglement (entanglement fidelity), which are closely related
\cite{Barnum KN 98}. For the present setup, suppose there is an additional
control spin $C$ which does not couple to the rest of the system,
but is initially maximally entangled with $S$, see figure \ref{entfig}.
A possible initial state is
\begin{equation}
\ve \Psi (0) \ket = (\ve \uparrow_C,
\uparrow_S, \downarrow_R, 0 \ket + \ve \downarrow_C, \downarrow_S,
\downarrow_R, 0 \ket)/\sqrt{2} \, .
\end{equation}
The transfer of the entanglement across the chain may now be analysed by considering the
entanglement between $R$ and $C$ as a function of time. Since we assume\linebreak
$J_a \ll (B, J_x, J_y, J_z)$, the state $\ve \downarrow_C,
\downarrow_S, \downarrow_R, 0 \ket$ is by virtue of energy
conservation approximately stationary, while the evolution of $\ve
\uparrow_C, \uparrow_S, \downarrow_R, 0 \ket$ is the same as above
(modulo a phase). Since states with more than one excitation are
energetically not accessible, the logarithmic negativity
\cite{Ple05} for the reduced density matrix of $R$ and $C$ can be
expressed approximately as
\begin{equation}
E_N \approx \log_2 \lk( P(\downarrow_S \uparrow_R)(t) + 1 \rk) \, .
\end{equation}
Hence, in this model $P(\downarrow_S \uparrow_R) \approx 1$
implies that the entanglement has been transferred perfectly, too.

\section{Conclusions}

In conclusion, we have considered entanglement and excitation transfer
through strongly coupled quantum many body systems.
In particular we have studied the dependence of the transfer
quality and speed on the size of the spectral gap between the ground
and the lowest excited state of the considered system.

As a first main result, we find that the quality of transfer,
and hence the quantum channel capacity, can be almost perfect
whenever there is a finite, sufficiently large energy gap above the ground state.
This opens up a generic way to design good quantum channels by using
gapped systems, since the gap ensures high transfer quality irrespective of the
system's details.

On approaching quantum critical points, the spectral gap shrinks and the transfer
decreases in quality, but accelerates.
This second main result suggests a possible experimental determination
of the energy gap: if one
finds in an experiment that the energy is not completely transferred
from one ancilla to the other, one can infer that the energy gap
is smaller than the available energy. The bound can be made
tighter by lowering the energy available in the ``sender''
ancilla. This procedure might in particular be helpful for
cold atom systems in optical lattices, where standard spectroscopy
is not applicable.

A quantitative study of the scaling of the transfer quality and
time in between the two detected scenarios and its relation to
critical exponents of various quantum phase transition universality classes
\cite{Sachdev1999} should be a subject of future research.
In that way an approach that originated in
quantum information considerations might open up a new way to
characterise and experimentally detect quantum phase transitions.

\section{Acknowledgements}
The authors would like to thank Sougato Bose and Daniel Burgarth
for discussions at early stages of this project. This work is part
of the QIP-IRC supported by EPSRC (GR/S82176/0), the Alexander von
Humboldt Foundation, Hewlett-Packard and the EU Integrated Project
QAP.


%

\end{document}